\documentclass[pre,preprint,showpacs,preprintnumbers,amsmath,amssymb]{revtex4}

\usepackage{graphicx}
\usepackage{bm}

\usepackage{amsfonts}

\begin{document}

\title{Stable three-dimensional modon soliton in plasmas}
\author{Volodymyr M. Lashkin}
\email{vlashkin@ukr.net}
 \affiliation{Institute for Nuclear
Research, Kiev 03680, Ukraine}

\date{\today}

\begin{abstract}
We derive the nonlinear equations that describe coupled drift
waves and ion acoustic waves in a plasma. We show that when the
coupling to ion acoustic waves is negligible, the reduced
nonlinear equation is a generalization of the Hasegawa-Mima
equation to the three-dimensional (3D) case. We find an exact
analytical solution of this equation in the form 3D soliton drift
wave (3D modon). By numerical simulations we study collisions
between the modons and show that the collisions can be fully
elastic.
\end{abstract}


\maketitle

\section{Introduction}
\label{sec1}

Localized coherent structures, such as solitons and vortices, are
universal objects which appear in many nonlinear physical systems
and, in particular, in laboratory and space plasmas
\cite{Petviashvili,Horton_Ichikawa}. In the dissipationless limit
the basic model equation for two-dimensional (2D) nonlinear drift
plasma waves is the Hasegawa-Mima (H-M) equation
\cite{Hasegawa_Mima78}. The same equation describes nonlinear
barotropic Rossby waves in atmospheres of rotating planets and in
oceans, and it is known as the Charney equation \cite{Charney}.
Larichev and Reznik found an exact analytic solution of the
Charney equation in the form of the 2D solitary dipole vortex (the
modon) \cite{Larichev76}. A remarkable property of this solution
is the stability of modons under head-on and overtaking collisions
with zero impact parameter between the modons
\cite{Kamimura81,Zabusky}. In these cases the modons preserve
their form after the collisions, and they behave just like the
one-dimensional solitons in nonlinear Schr\"{o}dinger (NLS)
equation and the Korteweg-de Vries equation (KdV)
\cite{Zakharov,Ablowitz}. The modon solution for the H-M equation
was written in Ref. \cite{Kamimura81}. Furthermore, various
generalizations of the H-M equation to other branches of plasma
oscillations, including drift Alfv\'{e}n waves, drift-flute modes
etc., have been widely used for obtaining the 2D modon solutions
(see e.g., \cite{Petviashvili,Horton_Ichikawa} and references
therein).

In three-dimensional (3D) case, when it is necessary to take into
account the ion motion along the external magnetic field
($z$-direction), nonlinear equations for drift waves were obtained
in Ref. \cite{Meiss83}. In the dissipationless limit these
equations have the 2D (pseudo 3D) solution of the form
Larichev-Reznik modon. In geophysics, 3D Rossby waves correspond
to the baroclinic model with vertical wave motion in an atmosphere
\cite{Pedlosky}. Three-dimensional baroclinic modon solution was
found by Berestov \cite{Berestov1,Berestov2}. Under this, the
solution was restricted to the region of vertical coordinate
$z\geqslant 0$.

The aim of the present work is to derive nonlinear equations
governing the dynamics of coupled drift and ion acoustic waves
without the assumption of plasma quasineutrality, to reduce these
equations to 3D analog of the H-M equation and to obtain exact
analytic 3D modon solutions. Note that, in contrast to Refs.
\cite{Berestov1,Berestov2} , the solutions do not assume
$z\geqslant 0$ and the hard lid boundary condition at $z=0$, and,
thus, include the antisymmetrical $z$-component.

The paper is organized as follows. In Sec. II, we derive the model
equations for coupled drift and ion acoustic plasma waves. The 3D
modon solutions are found in Sec. III. The collisions between the
modons are studied numerically in Sec. IV. Finally, a brief
summary of the results is presented in Sec. V.

\section{Model equations}
\label{sec2}

 We consider a plasma of cold ions and massless electrons (i.e.
 the characteristic frequency of motion much less than the electron
 plasma frequency)
in a homogeneous external magnetic field
$\mathbf{B}_{0}=B_{0}\mathbf{\hat{z}}$, where $\mathbf{\hat{z}}$
is the unit vector along the $z$-direction. The plasma motions are
described by the system of fluid equations for ions
\begin{equation}
\label{continuity_eq} \frac{\partial n_{i}}{\partial
t}+\nabla\cdot (n_{i}\mathbf{v})=0
\end{equation}
\begin{equation}
\label{motion_eq} \frac{\partial \mathbf{v}}{\partial
t}+\left(\mathbf{v}\cdot\nabla
\mathbf{v}\right)=-\frac{e}{M}\nabla\varphi+\Omega_{i}[\mathbf{v}\times
\hat{\mathbf{z}}]
\end{equation}
and the Poisson equation
\begin{equation}
\label{poisson} \Delta\varphi=4\pi e(\delta n_{e}-\delta n_{i}),
\end{equation}
where $n_{i}=n_{0}(x)+\delta n_{i}$ is the ion plasma density,
$n_{0}(x)=n_{0}(1+x/L)$ is the inhomogeneous equilibrium plasma
density, $L=(\partial_{x} \ln n_{0}(x))^{-1}$ is the logarithmic
equilibrium density length scale, $\mathbf{v}$ is the ion
velocity, $\varphi$ is the electrostatic potential, $\delta n_{e}$
and $\delta n_{i}$ are the electron and ion plasma density
perturbations respectively,  $M$ is the ion mass,
$\Omega_{i}=eB_{0}/Mc$ is the ion gyrofrequency. For the massless
electrons we assume Boltzmann distribution $\delta
n_{e}=e\varphi/T_{e}$, where $T_{e}$ is the electron temperature.
We note that the Boltzmann distribution for electrons does not
imply the quasineutrality and is violated on the scale length of
the Debye radius.

We assume that temporal variation of perturbations is slow
compared to the frequency of ion gyrations
\begin{equation}
\label{small} \frac{\partial /\partial t}{\Omega_{i}}\sim
\frac{(\mathbf{v}\cdot\nabla )}{\Omega_{i}}\ll 1 .
\end{equation}
Using Eq. (\ref{motion_eq}), with the accuracy to the first order
in small parameter Eq. (\ref{small}) , we can write the ion
velocity perpendicular to the magnetic field as
\begin{equation}
\label{v_bot}
\mathbf{v}_{\bot}=\mathbf{v}_{E}-\frac{e}{M\Omega_{i}^{2}}\frac{d}{d
t}\nabla_{\bot}\varphi,
\end{equation}
where we have introduced the notation $\mathbf{v}_{E}$ for the
$\mathbf{E}\times \mathbf{B}$ drift velocity
\begin{equation}
\mathbf{v}_{E}=\frac{e}{M\Omega_{i}}[ \hat{\mathbf{z}}\times
\nabla_{\bot}\varphi],
\end{equation}
and
\begin{equation}
\frac{d}{dt}=\frac{\partial}{\partial t}+(\mathbf{v}_{E}\cdot
\nabla_{\bot})\equiv \frac{\partial}{\partial
t}+\frac{e}{M\Omega_{i}}\{\varphi , \cdots \},
\end{equation}
where the Poisson bracket is defined by
\begin{equation}
\label{Pois_brack} \{f,g\}=\frac{\partial f}{\partial
x}\frac{\partial g}{\partial y}-\frac{\partial f}{\partial
y}\frac{\partial g}{\partial x}.
\end{equation}
The main nonlinearity in Eqs. (\ref{continuity_eq}) and
(\ref{motion_eq}) comes from the $\mathbf{v}_{E}$ convection term.
From of Eq. (\ref{continuity_eq}) we have
\begin{equation}
\label{continuity} \frac{\partial \delta n_{i}}{\partial
t}+n_{0}\nabla\cdot
(\mathbf{v}_{\bot}+\hat{\mathbf{z}}v_{z})+\mathbf{v}_{E}\cdot\nabla
\delta n_{i}+\frac{e n_{0}}{M
\Omega_{i}L}\frac{\partial\varphi}{\partial y}=0.
\end{equation}
Substituting $\mathbf{v}_{\bot}$ from Eq. (\ref{v_bot}) into Eq.
(\ref{continuity}) , and eliminating $\delta n_{i}$ with the aid
of Eq. (\ref{poisson}), we get
\begin{gather}
\label{eq1}
 \frac{\partial}{\partial t}\left(\varphi-\rho_{s}^{2}\Delta_{\bot}\varphi
 -R^{2}\Delta\varphi\right)+v_{d}\frac{\partial\varphi}{\partial
 y}+\frac{Mc_{s}^{2}}{e}\frac{\partial v_{z}}{\partial
 z}
\\
 -\frac{e}{M\Omega_{i}}\left\{\varphi,\rho_{s}^{2}\Delta_{\bot}\varphi
 +R^{2}\Delta\varphi\right\}=0
\end{gather}
where $c_{s}=\sqrt{T_{e}/M}$ is the ion sound velocity,
$v_{d}=c_{s}^{2}/\Omega_{i}L$ is the drift velocity,
$\omega_{pi}=\sqrt{4\pi e^{2}n_{0}/M}$ is the ion plasma
frequency, $R=c_{s}/\omega_{pi}$ is the Debye length,
$\rho_{s}=c_{s}/\Omega_{i}$. Taking into account that
$\Omega_{i}/\omega_{pi}\ll 1$, one can obtain
\begin{gather}
\label{eq2}
 \frac{\partial}{\partial t}\left(\varphi-\rho_{s}^{2}\Delta_{\bot}\varphi
 -R^{2}\frac{\partial^{2}\varphi}{\partial z^{2}}\right)+v_{d}\frac{\partial\varphi}{\partial
 y}+\frac{Mc_{s}^{2}}{e}\frac{\partial v_{z}}{\partial
 z} \\ \nonumber
 -\frac{e}{M\Omega_{i}}\left\{\varphi,\rho_{s}^{2}\Delta_{\bot}\varphi
 +R^{2}\frac{\partial^{2}\varphi}{\partial z^{2}}\right\}=0
\end{gather}
From $z$-component of Eq. (\ref{motion_eq}) we have
\begin{equation}
\label{motion_z_eq} \frac{\partial v_{z}}{\partial
t}+(\mathbf{v}_{E}\cdot
\nabla_{\bot})v_{z}=-\frac{e}{M}\frac{\partial \varphi}{\partial
z}.
\end{equation}
In the linear approximation, taking $\varphi\sim \exp
(-i\mathbf{k}\cdot \mathbf{r}+i\omega t)$, Eqs. (\ref{eq2}) and
(\ref{motion_z_eq}) yield the dispersion relation
\begin{equation}
\label{disp_my} \omega^{2}\left(1+k_{\bot} ^{2} \rho_{s}
^{2}+k_{z}^{2}R^{2}\right)-\omega k_{y}v_{d}-k_{z}^{2}c_{s}^{2}=0
\end{equation}
with $k_{\bot}^{2}=k_{x}^{2}+k_{y}^{2}$ , where $k_{x}$, $k_{y}$
and $k_{z}$ are the components of the wave vector $\mathbf{k}$.
From Eq. (\ref{disp_my})  it follows that there are two branches
of plasma oscillations: the drift wave ( if $\omega k_{y}v_{d}\gg
k_{z}^{2}c_{s}^{2}$)
\begin{equation}
\label{disp_drift} \omega=\frac{k_{y}v_{d}}{1+k_{\bot} ^{2}
\rho_{s} ^{2}+k_{z}^{2}R^{2}}
\end{equation}
and the ion acoustic wave (if $\omega k_{y}v_{d}\ll
k_{z}^{2}c_{s}^{2}$)
\begin{equation}
\label{disp_sound} \omega^{2}=\frac{k_{z}^{2}c_{s}^{2}}{1+k_{\bot}
^{2} \rho_{s} ^{2}+k_{z}^{2}R^{2}}
\end{equation}
Neglecting the interaction with the ion sound (i.e. $k_{z}c_{s}\ll
\omega\sim k_{y}v_{d}$) in Eqs. (\ref{eq2}) and
(\ref{motion_z_eq}) (physically, this assumption means that the
ion inertia in the direction of the ambient magnetic field is
negligible),  we obtain the equation for the potential $\varphi$
\begin{equation}
\label{eqbasic}
 \frac{\partial}{\partial t}\left(\varphi-\rho_{s}^{2}\Delta_{\bot}\varphi
 -R^{2}\frac{\partial^{2}\varphi}{\partial z^{2}}\right)+v_{d}\frac{\partial\varphi}{\partial
 y}
  -\frac{e}{M\Omega_{i}}\left\{\varphi,\rho_{s}^{2}\Delta_{\bot}\varphi
 +R^{2}\frac{\partial^{2}\varphi}{\partial z^{2}}\right\}=0
\end{equation}
In the following we introduce the dimensionless variables
\begin{gather}
\label{dimensionless} \Phi\rightarrow e\varphi/T_{e},
\,\,\mathbf{r}_{\bot}\rightarrow \mathbf{r}_{\bot}/\rho_{s}, \,\,
z\rightarrow z/R, \\ \nonumber t\rightarrow \Omega_{i}t, \,\,
v_{d}\rightarrow v_{d}/c_{s}.
\end{gather}
Note here that we use the stretched dimensionless $z$ coordinate.
Equation (\ref{eqbasic}) then becomes
\begin{equation}
\label{basic_eq1} \frac{\partial}{\partial
t}(\Phi-\Delta\Phi)+v_{d}\frac{\partial\Phi}{\partial
y}-\{\Phi,\Delta\Phi\}=0
\end{equation}
and Eq. (\ref{basic_eq1}) can be  rewritten as
\begin{equation}
\label{basic_eq2} \frac{\partial \Gamma}{\partial
t}+\{\Phi,\Gamma\}=0
\end{equation}
where $\Gamma=\Delta\Phi-\Phi+v_{d}x$ is the generalized
vorticity. Equation (\ref{basic_eq2}) describes the convection of
the generalized vorticity with $d\Gamma/dt=0$ and has an infinite
set of integrals of motion (Casimir invariants)
\begin{equation}
\label{integrals} \int f(\Gamma,z)\,d\mathbf{r},
\end{equation}
where $f$ is an arbitrary function of its arguments. Other
integrals of motion are
\begin{equation}
\label{integrals}  \int \Phi\Gamma\,d\mathbf{r} \, ,  \, \int
\,x\Gamma\, d\mathbf{r} \, ,  \, \int \,(y-v_{d}t)\Gamma\,
d\mathbf{r}.
\end{equation}
In particularly, the quadratic invariants for the energy $E$ and
the enstrophy $K$ are
\begin{gather}
\label{integrals_E_K}  E=\int
\left[\Phi^{2}+(\nabla\Phi)^{2}\right]\,d\mathbf{r}   , \\
K=\int
\left[(\nabla\Phi)^{2}+(\Delta\Phi)^{2}\right]\,d\mathbf{r}.
\end{gather}
The existence of an infinite number of integrals of motion might
suggest that the system is complete integrable. However, as was
shown in \cite{Shulman88} for 2D Charney-Hasegava-Mima equation,
the existence of infinitely many invariants other than the
Casimirs does not imply the integrability and, in particularly,
additional constraints on the wave spectrum are required.

\section{3D modon soliton solution}
\label{sec3}

We look for stationary traveling wave solutions of Eq.
(\ref{basic_eq2}) of the form
\begin{equation}
\label{mov_reference} \Phi(x,y,z,t)=\Phi(x,y',z), \, \, \,
y'=y-ut,
\end{equation}
where $u$ is the velocity of propagation in the $y$ direction.
Substituting Eq. (\ref{mov_reference}) into Eq. (\ref{basic_eq2})
yields a nonlinear equation for $\Phi(x,y',z)$ (in the following
we omit the prime), which can be written as a single Poisson
bracket relation
\begin{equation}
\label{stat_forma} \left\{\Gamma,\Phi-ux\right\}=0.
\end{equation}
This implies that the functions in the bracket are dependent and,
therefore, we have
\begin{equation}
\label{gen_solution} \Gamma=F(\Phi-ux,z)
\end{equation}
where $F$ is an arbitrary function of its arguments. Following the
known procedure for finding 2D modon solutions
\cite{Larichev76,Petviashvili,Horton_Ichikawa}, we consider two
forms of the function $F$, namely $F_{int}$ and $F_{ext}$ - for
the interior region and the exterior region of 3D space.
 The boundary between the region containing the streamlines,i.e.
 the isolines of the function
$\Phi-ux$, which
 extend to infinity and the region containing the streamlines which do
 not extend to infinity is assumed to be the sphere
 $r\equiv\sqrt{x^{2}+y^{2}+z^{2}}=a$, where the parameter $a$ is
the modon radius. We are looking for localized solutions, that is
$\Phi\rightarrow 0$ as $x,y,z\rightarrow \pm\infty$, i. e. for the
streamlines which extend to infinity. Considering the limit
$x,y,z\rightarrow \pm\infty$ of Eq. (\ref{gen_solution}), one can
conclude that the function $F_{ext}$ must be linear for the
localized solutions, that is
\begin{equation}
\label{eq_ext}
 \Delta\Phi-\Phi+v_{d}x=c_{1}(\Phi-ux)+c_{2}+c_{3}z
\end{equation}
with $c_{1}=-v_{d}/u$, $c_{2}=0$,$c_{3}=0$. One can see that we
must have $1-v_{d}/u>0$  for the solution to be localized, that
is, $u<0$ or $u>v_{d}$. Thus, the modon velocity $u$ must be
outside the region of possible phase velocities $\omega/k_{y}$ of
the linear drift wave solution of Eq. (\ref{eqbasic}).

For those streamlines that do not extend to infinity there is no
boundary condition to  \textit{a priori} determine the form of the
function $F$ in the interior region. The simplest choice of
$F_{int}$ is also the linear function and we have
\begin{equation}
\label{eq_int}
 \Delta\Phi-\Phi+v_{d}x=c_{4}(\Phi-ux)+c_{5}+c_{6}z
\end{equation}
where $c_{4}$, $c_{5}$ and $c_{6}$ are arbitrary. The requirement
of finiteness at $r=0$ implies $1+c_{4}>0$. In the following we
introduce the notatons
\begin{equation}
\label{beta_gamma} \varkappa=a\sqrt{1-v_{d}/u}, \,
k=a\sqrt{1+c_{4}}.
\end{equation}
Equation (\ref{eq_ext}) has a  solution
\begin{equation}
\sum_{n,l,m}A_{nlm}\frac{K_{n+1/2}(\varkappa r/a)}{\sqrt{
r}}Y_{lm}(\theta,\varphi),
\end{equation}
while a solution of Eq. (\ref{eq_int}) is
\begin{equation}
\sum_{n,l,m}B_{nlm}\frac{J_{n+1/2}(k r/a)}{\sqrt{
r}}Y_{lm}(\theta,\varphi),
\end{equation}
where we use spherical coordinates $(r,\theta,\varphi)$, $n,m,l$
are integers, $J_{\nu}(\xi)$ is the Bessel function of the first
kind, $K_{\nu}(\xi)$ is the modified Bessel function of the second
kind, $Y_{lm}$  are the spherical harmonics, $A_{nlm}$ and
$B_{nlm}$ are arbitrary constants. At present we consider only the
lowest modes in these sums consistent with the terms $v_{d}x$ and
$c_{6}z$, namely $n=0,1$, $l=0,1$, and $m=0,1$. We require that
$\Phi$ and $\nabla\Phi$ to be continues at $r=a$
\begin{equation}
\Phi\mid_{r=a-0}=\Phi\mid_{r=a+0}, \, \,
\nabla\Phi\mid_{r=a-0}=\nabla\Phi\mid_{r=a+0},
\end{equation}
and $\Delta\Phi$ (or, equivalently, $\Gamma$) has a constant jump
$p$ (including the case $p=0$) at $r=a$
\begin{equation}
\Delta\Phi\mid_{r=a-0}=\Delta\Phi\mid_{r=a+0}+p.
\end{equation}
Then, for a given value of $\beta$, the value of $k$ is determined
by the relation
\begin{equation}
\label{transzent} (\delta k^{2}+3-k^{2})\tan k=k(\delta k^{2}+3)
\, ,
\end{equation}
where
\begin{equation}
\label{delta}
\delta=\frac{(\varkappa^{2}+3\varkappa+3)}{\varkappa^{2}(\varkappa+1)}.
\end{equation}
The final solution is
\begin{equation}
\label{final_solution}
\Phi(r,\theta,\varphi)=\Psi_{0}(r)+\Psi(r)(\sin\theta\cos\varphi+\mu\cos\theta),
\end{equation}
where
\begin{widetext}
\begin{equation}
\label{psi0}
\Psi_{0}(r)=\frac{pa^{2}}{(\varkappa^{2}+k^{2})\delta}\left\{
\begin{array}{lc}
\displaystyle  \, \frac{a\sin (k r/a)}{r(\sin k-k\cos
k)}-\frac{3(\varkappa^{2}+k^{2})}{\varkappa^{2}k^{2}},
&  r\leqslant a \\
\displaystyle \,\frac{a}{(1+\varkappa)r}\,e^{-\varkappa (r/a
-1)},& r\geqslant a
\end{array}
\right. ,
\end{equation}

\begin{equation}
\label{psi} \Psi(r)=ua\left\{
\begin{array}{lc}
\displaystyle
\left(1+\frac{\varkappa^{2}}{k^{2}}\right)\frac{r}{a}-
\frac{\varkappa^{2}}{k^{2}}\frac{a^{2}[\sin (k
r/a)-(kr/a)\cos(kr/a)]}{r^{2}(\sin k-k\cos k)}
,&  r\leqslant a \\
\displaystyle  \,\frac{a^{2}(1+\varkappa
r/a)}{r^{2}(1+\varkappa)}\,e^{-\varkappa (r/a -1)},& r\geqslant a
\end{array}
\right. ,
\end{equation}
\end{widetext}
The solution is the sum of three terms: radially symmetric,
antisymmetric in the $x$-direction, and antisymmetric in the
$z$-direction. The radially symmetric part vanishes if $p=0$, and,
under this, $\Delta\Phi$ (and the vorticity) is continuous at the
boundary $r=a$. The $z$-antisymmetric part vanishes if $\mu=0$.
Thus, the modon solution (\ref{final_solution}) has four
independent free parameters - the velocity $u$, the modon radius
 $a$, the amplitude of the $z$-antisymmetric part $\mu$, and
the jump of the vorticity $p$. Within the interior region $r<a$,
the fluid particles are trapped and are thus transported along the
$y$-direction. In the exterior region $r>a$, the solution decays
exponentially to zero. In the limiting case $\varkappa\rightarrow
0$, that is $u\rightarrow v_{d}$, we have
\begin{widetext}
\begin{equation}
\label{psi0_bet0} \Psi_{0}(r)=\frac{pa^{2}}{k^{2}}\left\{
\begin{array}{lc}
\displaystyle \, \frac{a\sin (kr/a)}{r\sin k}-1,
&  r\leqslant a \\
\displaystyle \,0,& r\geqslant a
\end{array}
\right. ,
\end{equation}

\begin{equation}
\label{psi_bet0} \Psi(r)=v_{d}a\left\{
\begin{array}{lc}
\displaystyle \frac{r}{a}- \frac{3a^{2}[\sin (k r/a)-(k r/a)\cos(k
r/a)]}{r^{2}k^{2}\sin k}
,&  r\leqslant a \\
\displaystyle  \,\frac{a^{2}}{r^{2}},& r\geqslant a
\end{array}
\right. ,
\end{equation}
\end{widetext}
In the other limiting case $\varkappa\rightarrow \infty$, that is
$u\rightarrow 0$,  we have for the radially symmetric component $
\Psi_{0}(r)=0$ and
\begin{widetext}
\begin{equation}
\label{psi_betinft} \Psi(r)=\frac{v_{d}a^{3}}{k^{2}}\left\{
\begin{array}{lc}
\displaystyle \frac{a^{2}[\sin (k r/a)-(k r/a)\cos(k
r/a)]}{r^{2}(\sin k-k\cos k)}-\frac{r}{a}
,&  r\leqslant a \\
\displaystyle  \,0,& r\geqslant a
\end{array}
\right. ,
\end{equation}
\end{widetext}
Equation (\ref{transzent}) has an infinite set of roots $k_{n}$,
$n=1,2\dots$ for each $\varkappa$. Therefore, Eqs.
(\ref{final_solution}), (\ref{psi0}) and (\ref{psi}) present the
infinite set of solutions with $k=k_{n}$. The solution with $n=1$
(the ground state modon) has no radial nodes. The higher states
have $n-1$ nodes (in the interior region). The functions
$\Psi_{0}(r)$ and $\Psi(r)$ for the lowest states with $n=1,2,3$
are plotted in Fig 1.
\begin{figure}
\includegraphics[width=6.8in]{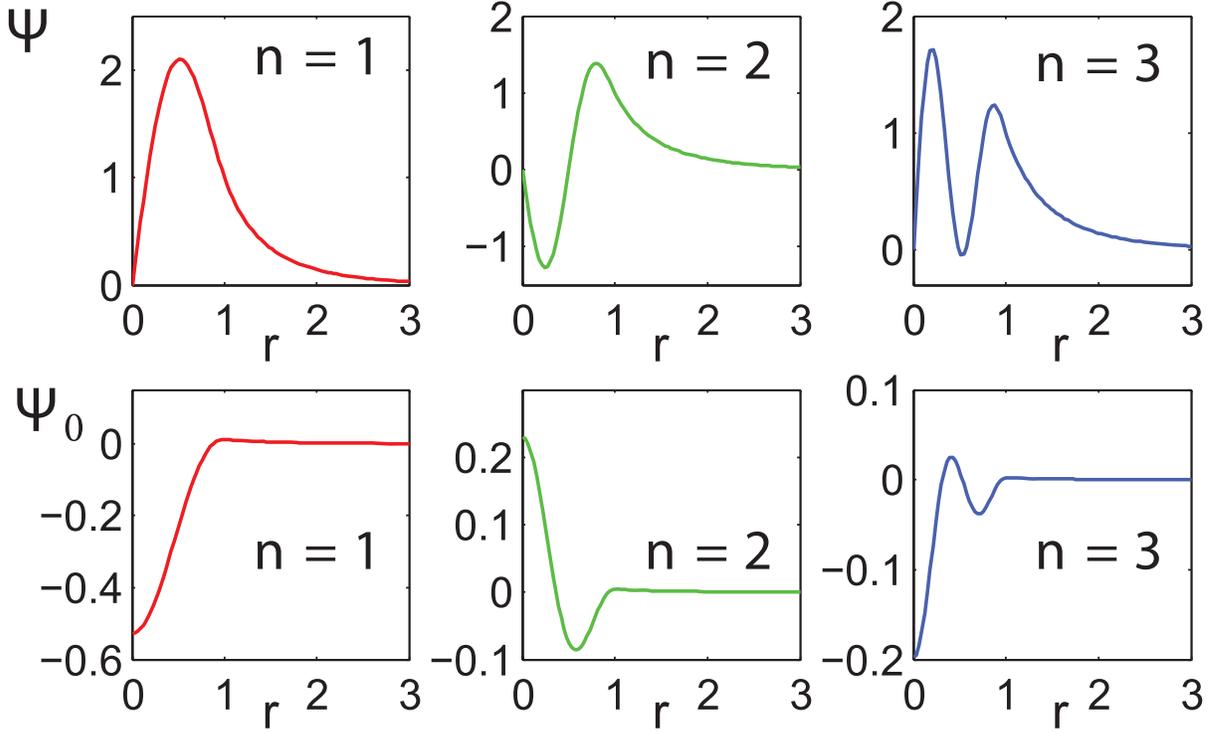}
\caption{\label{fig1}  (Color online)  Top row: the function
$\Psi$ as a function of the radial coordinate for the three lowest
states with $n=1,2,3$. The modon  parameters are $v_{d}=0.1$,
$a=1.$, $u=1.$, $p=2.$. Bottom row: the same for the function
$\Psi_{0}$. }
\end{figure}
Further we consider only the ground state modon with $k=k_{1}$.

The modon energy $E$ and enstrophy $K$ can be computed
straightforwardly
\begin{equation}
E=E_{0}+E_{1} , \, \, K=K_{0}+K_{1},
\end{equation}
where $E_{0}$ and $K_{0}$ are the energy and enstrophy of the
radially symmetric part respectively
\begin{equation}
\label{E0} E_{0}=\frac{2\pi
p^{2}a^{5}}{\delta^{2}\varkappa^{2}k^{2}(\varkappa^{2}+k^{2})}
\left[\delta^{2}\varkappa^{2}k^{2}+a^{2}(5\delta+1)\right] ,
\end{equation}
\begin{gather}
\label{K0} K_{0}=\frac{4\pi
p^{2}a^{3}}{\delta^{2}(\varkappa^{2}+k^{2})^{2}}
\left\{\frac{\delta^{2}}{2}\left[a^{2}(k^{2}-\varkappa^{2})+\varkappa^{4}+
k^{4}\right] \right. \\ \nonumber \left.
+(5\delta+1)\left(a^{2}-\frac{\varkappa^{2}}{2}
-\frac{k^{2}}{2}\right)+6\left(1-\frac{a^{2}}{\varkappa^{2}}\right)\right\},
 \end{gather}
and
\begin{equation}
\label{E}
E_{1}=\frac{4\pi}{15}u^{2}a^{5}(1+\mu^{2})\left(1+\frac{\varkappa^{2}}{k^{2}}\right)
\left[\delta^{2}\varkappa^{2}
+(5\delta+1)\left(\frac{7\varkappa^{2}}{5k^{2}}+\frac{2}{5}
-\frac{v_{d}}{u}\right)\right] ,
\end{equation}
\begin{equation}
\label{K1}
K_{1}=\frac{2\pi}{3}u^{2}a^{3}(1+\mu^{2})\varkappa^{2}\left(1+\frac{\varkappa^{2}}{k^{2}}\right)
\left[\left(\frac{\delta\varkappa
k}{a}\right)^{2}+5\delta+1\right] ,
\end{equation}

In dimensional variables  (\ref{dimensionless}) one can see that
the 3D modon structure is flattened in a plane perpendicular to
the direction of the external magnetic field ($xy$ plane) with
asymmetry factor $\rho_{s}/R\equiv\omega_{pi}/\Omega_{i}\gg 1$.

\section{Collisions between modons}
\label{sec4}

In this section, we study the time evolution of the modons under
their collisions. To this end, we numerically solve the nonlinear
equation (\ref{basic_eq1}) with the initial conditions given by a
superposition of exact analytical modon solutions
(\ref{final_solution}),i.e.
\begin{equation}
\Phi(\mathbf{r},t)=\Phi_{mod}(\mathbf{r}-\mathbf{r}_{1},t)+\Phi_{mod}(\mathbf{r}-\mathbf{r}_{2},t)
\end{equation}
 at the time $t=0$. The time integration is performed
by an  implicit Adams-Moulton method with the variable timestep
and the variable order, and local error control. The periodic
boundary conditions are assumed. The linear terms are computed in
spectral space. The Poisson bracket nonlinearity is evaluated in
physical space by a finite difference method, using the energy and
enstrophy conserving Arakawa scheme \cite{Arakawa}. Total energy
and enstrophy were conserved with a relative accuracy less than
$5\times 10^{-4}$ during the simulations.

As a first case we consider a collision between the modons which
move in opposite directions along the $y$-axis (the head-on
collision). The collision is assumed to be with the zero impact
parameter, i.e. $x_{1}=x_{2}$, $z_{1}=z_{2}$. The parameter
$v_{d}$ is set to $v_{d}=0.1$ in all simulations. The modon
velocities are $u_{1}=0.3$ and $u_{2}=-0.5$, and the modon
radiuses are $a_{1}=a_{2}=0.5$. The modons are considered to be
without radially symmetric and $z$-antisymmetric components, that
is $p_{1}=p_{2}=\mu_{1}=\mu_{2}=0$. The time evolution of modons
is presented in Fig. 2.
\begin{figure}
\includegraphics[width=6.8in]{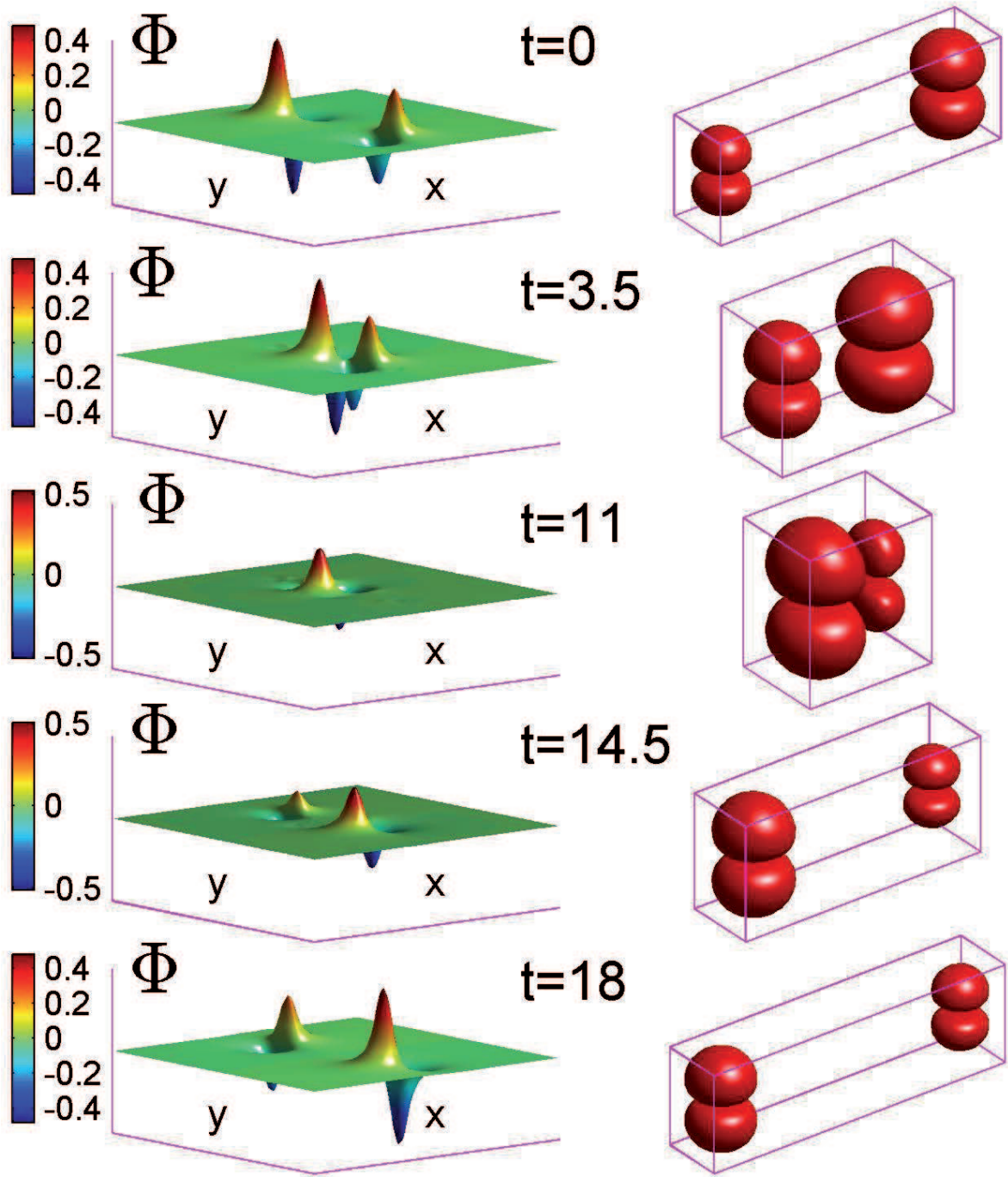}
\caption{\label{fig2} (Color online) Elastic head-on collision
between the modons. The modon parameters are $v_{1}=0.3$,
$v_{2}=-0.5$, $a_{1}=0.5$, $a_{2}=0.5$,
$p_{1}=p_{2}=\mu_{1}=\mu_{2}=0.$. The left column: the field
distribution $\Phi$ in the $x-y$ plane. The right column: the
isosurface $\Phi(x,y,z)=0.15$.}
\end{figure}
The modons approach each other, $t=3.5$, and undergo a complicated
interaction with strong overlapping during the collision, $t=11$,
thus generating strong mutual disturbances, as shown at various
time intervals in Fig. 2. They then pass through each other and
begin to separate, $t=14.5$, and, after all, fully reconstructing
their initial form without any emitting wakes of radiation,
$t=18$.

The second example is  a collision when two modons (without
radially symmetric and $z$-antisymmetric components) move in the
same direction along the $y$-axis (the overtaking collision). The
zero impact parameter collision  is also assumed. The modon
velocities are essentially different with $u_{1}=-0.3$,
$u_{2}=-0.01$ and the modon radiuses are different with
$a_{1}=0.5$, $a_{2}=2$. The time evolution is presented in Fig. 3.
\begin{figure}
\includegraphics[width=6.8in]{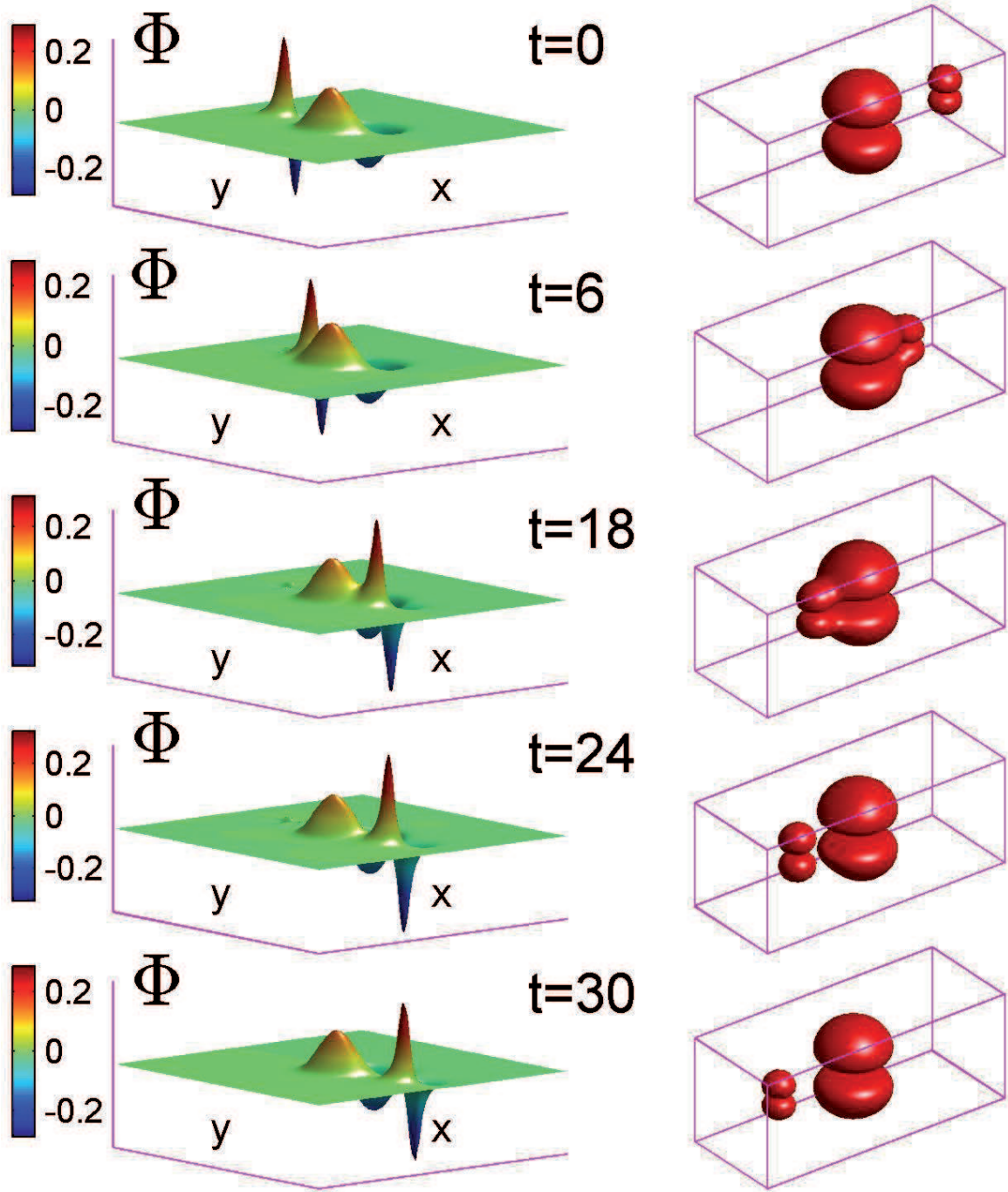}
\caption{\label{fig3} (Color online) Elastic  overtaking collision
between the modons. The modon parameters are $v_{1}=-0.3$,
$v_{2}=-0.01$, $a_{1}=0.5$, $a_{2}=2.$,
$p_{1}=p_{2}=\mu_{1}=\mu_{2}=0.$. The left column: the field
distribution $\Phi$ in the $x-y$ plane. The right column:
isosurface $\Phi(x,y,z)=0.08$. }
\end{figure}
The fast modon catches up with the the slow one, $t=6$, then the
modons overlap and pass through each other,$t=18$,begin to
separate, $t=24$, and recover their shape, $t=30$. The simulations
have been performed for various values of modon velocities and
radiuses. Thus, for the 3D modons without radially symmetric and
$z$-antisymmetric parts the collisions are elastic for both the
head-on and overtaking cases (for the zero impact parameter). Such
behavior resembles collisions between one-dimensional solitons in
completely integrable models like the NLS and the KdV. The nonzero
impact parameter head-on collisions as well as overtaking ones
turn out to be inelastic. The modons are destroyed during the
course of collision, though for the small impact parameter
(compared to modon radiuses) the modons pass through each other
almost without changing their shape leaving wakes of radiation,
but after all, are destroyed.

Next, we address collisions between modons with radially symmetric
($p_{1}\neq 0$,$p_{2}\neq 0$) or/and $z$-antisymmetric
($\mu_{1}\neq 0$,$\mu_{2}\neq 0$) parts. In this connection, it is
necessary to stress that the nonlinearity in Eq. (\ref{basic_eq1})
is identically zero for any field distributions with only radial
symmetry or/and antisymmetry with the respect to the $z$-axis.
Hence, the terms with $p\neq 0$ and/or $\mu\neq 0$ in the solution
(\ref{final_solution}) should be considered as nonlinear
perturbations, though the corresponding amplitudes can be much
greater than the amplitude of the $x$-antisymmetric component. The
head-on collisions with zero impact parameter are considered. The
modon parameters are the same as in the case of modons without
radially symmetric and $z$-antisymmetric components.  Time
evolution of modons with radially symmetric parts,$p_{1}=2.$ and
$p_{2}=0.3$, is presented in Fig. 4.
\begin{figure}
\includegraphics[width=6.8in]{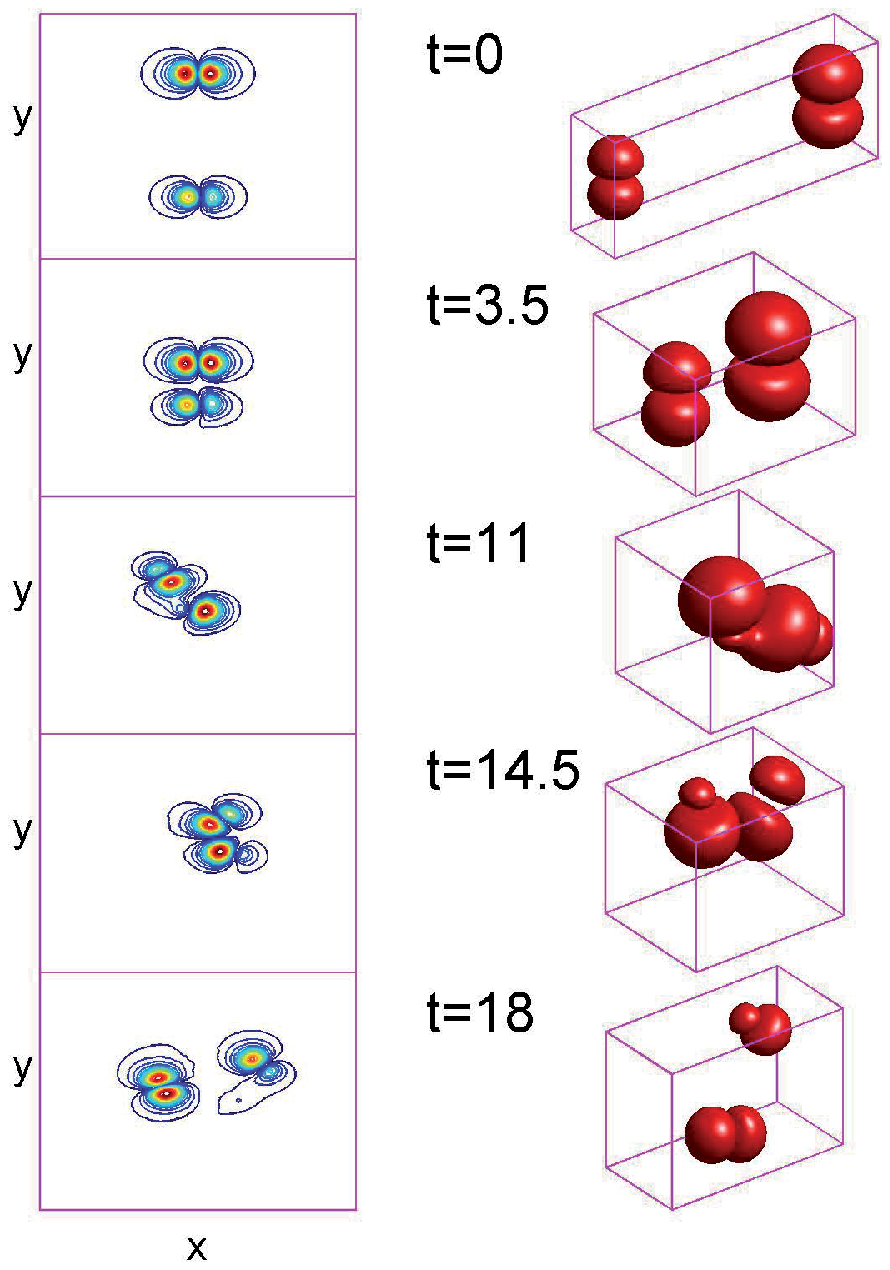}
\caption{\label{fig4}  (Color online) Inelastic head-on collision
of the modons with radially symmetric part. The modon parameters
are $v_{1}=0.3$, $v_{2}=-0.5$, $a_{1}=0.5$, $a_{2}=0.5$,
$p_{1}=2.$, $p_{2}=0.3$, $\mu_{1}=\mu_{2}=0$. The left column:
contour plots of $\Phi$  in the $x-y$ plane. The right column:
isosurfaces $\Phi(x,y,z)=0.2$.}
\end{figure}
It is seen that after approaching each other, $t=3.5$, and
overlapping, $t=11$, the modons propagate in a direction almost
perpendicular to the initial direction of propagation and are
destroyed during the course of collision, $t=14.5$ and $t=18$.

Inelastic collision of the modons with $z$-antisymmetric part,
$\mu_{1}=1.$ (i.e.,the amplitudes of $x$-antisymmetric and
$z$-antisymmetric parts are equal) and $\mu_{2}=0.5$ is shown in
Fig. 5.
\begin{figure}
\includegraphics[width=6.8in]{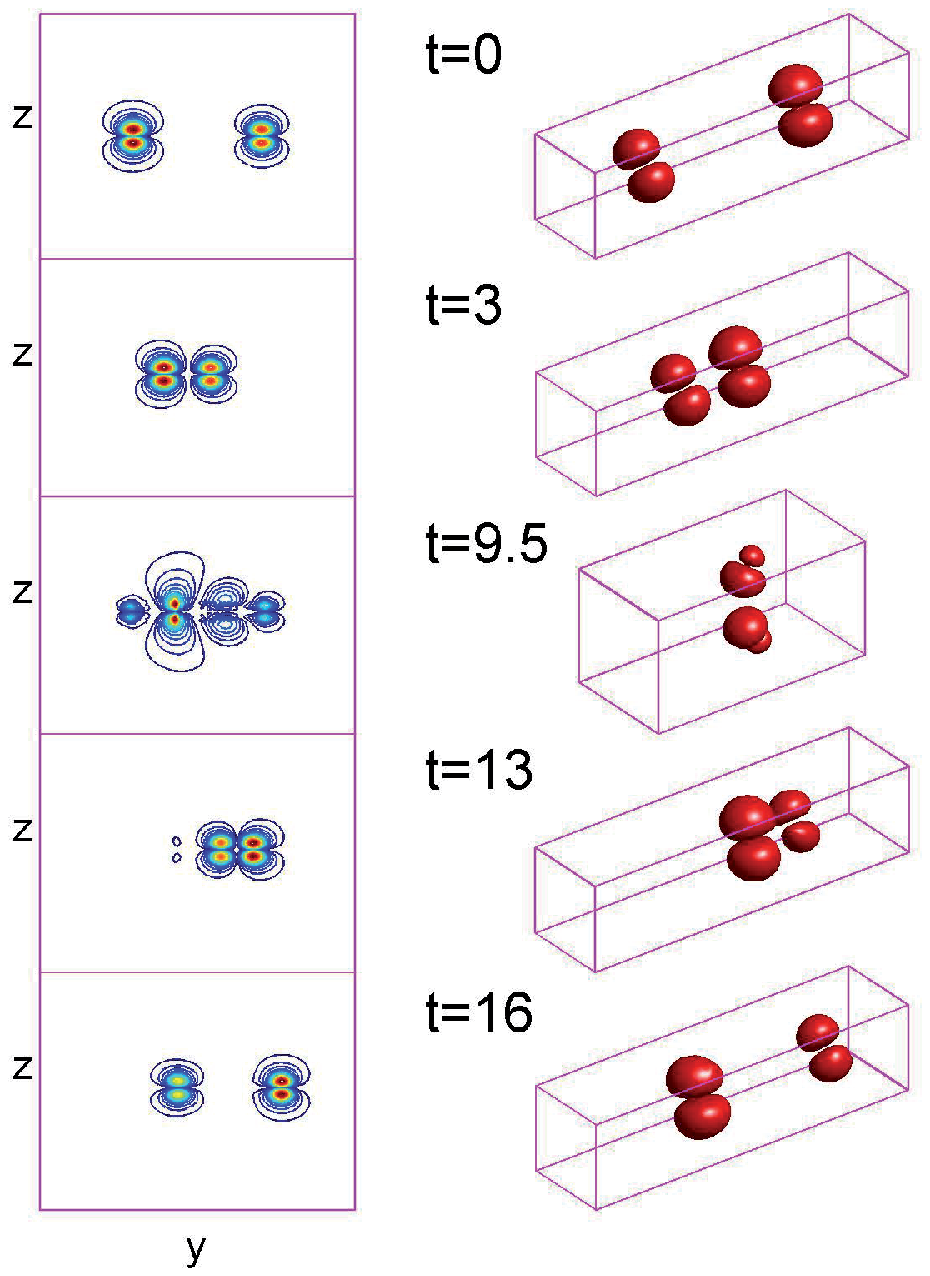}
\caption{\label{fig5}  (Color online) Inelastic head-on collision
of the modons with $z$-antisymmetric part.The amplitudes of modons
after the collison are changed due to emitted radiation. The modon
parameters are $v_{1}=0.3$, $v_{2}=-0.5$, $a_{1}=0.5$,
$a_{2}=0.5$, $p_{1}=p_{2}=0$, $\mu_{1}=1.$, $\mu_{2}=0.5$. The
left column: contour plots of $\Phi$ in the $y-z$ plane. The right
column: isosurfaces $\Phi(x,y,z)=0.2$. }
\end{figure}
In this case the modons almost preserve their initial shape after
collision, $t=16$, resulting in a slightly decreasing the modon
amplitudes, but are destroyed at large times (not shown here) due
to energy and enstrophy loss connected with emitted radiation.
Modons with sufficiently large amplitudes of $x$-antisymmetric
parts, $\mu_{1,2}\gg 1$, are fully destroyed immediately after the
collision.

\section{Discussion and conclusions}
\label{sec5}

We have derived the nonlinear equations that describe coupled
drift waves and  ion acoustic waves in a plasma assuming  electron
adiabaticity and negligible ion pressure. We have shown that when
the coupling to ion acoustic waves is negligible, the reduced
equation is a generalization of the H-M equation to the 3D case.
We have found an exact analytical solution of this equation in the
form 3D solitary nonlinear drift wave (3D modon). We have
performed numerical simulations to study the stability of modons
under collisions. The simulations show that the modons without
radially symmetric and $z$-antisymmetric parts preserve their
shape after the zero-impact parameter collisions (fully elastic
soliton collisions) and there is no emitted radiation. This is
true for both head-on and overtaking collisions. The modons with
radially symmetric or/and $z$-antisymmetric parts are  destroyed
after collisions. The nonzero-impact parameter collisions between
modons are inelastic.

The question of the  stability of modons with respect to arbitrary
perturbations is still open. The stability properties of the 2D
modons was investigated in Ref. \cite{Swaters} (the 2D modons with
$v>v_{d}$ are always unstable because of the tilt instability,
while the modons with $v<0$ may be stable for some region of
parameters). However, as is well known, the stability of soliton
structures depends strongly on the dimensionality of  space (see,
for example, Ref. \cite{Kivshar}). The stability of the 3D modons
will be discussed in a forthcoming publication.

In conclusion we note, that  Eq. (\ref{basic_eq1}) can be modified
straightforwardly by including an additional nonlinearity of the
KdV type $\sim \Phi
\partial\Phi/\partial y$ taking into account the effect of the
electron temperature gradient \cite{Su}.

\end{document}